\documentclass[fleqn,usenatbib]{mnras}

\usepackage[T1]{fontenc}
\usepackage[utf8]{inputenc}
\usepackage{multicol}
\usepackage{comment}

\DeclareRobustCommand{\VAN}[3]{#2}
\let\VANthebibliography\thebibliography
\def\thebibliography{\DeclareRobustCommand{\VAN}[3]{##3}\VANthebibliography}

\usepackage{graphicx}	
\usepackage{amssymb}	
\usepackage{newtxtext,newtxmath}

\title[]{Observability of Low-Luminosity AGN in the Early Universe with JWST}

\author[Jeon J. et al.]{
Junehyoung Jeon\textsuperscript{\href{https://orcid.org/0000-0002-6038-5016}{\includegraphics[width=2.5mm]{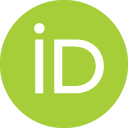}}\,}\thanks{E-mail: junehyoungjeon@utexas.edu}$^{1}$, 
Boyuan Liu\textsuperscript{\href{https://orcid.org/0000-0002-4966-7450}{\includegraphics[width=2.5mm]{orcid.png}}}$^{2}$ 
Volker Bromm\textsuperscript{\href{https://orcid.org/0000-0003-0212-2979}{\includegraphics[width=2.5mm]{orcid.png}}}$^{1}$ 
and Steven L.~Finkelstein\textsuperscript{\href{https://orcid.org/0000-0001-8519-1130}{\includegraphics[width=2.5mm]{orcid.png}}}$^{1}$
\\
$^{1}$Department of Astronomy, University of Texas, Austin, TX 78712, USA\\
$^{2}$University of Cambridge Institute of Astronomy: Cambridge, Cambridgeshire, GB\\
}

\date{}

\pubyear{2022}

\begin{document}
\label{firstpage}
\pagerange{\pageref{firstpage}--\pageref{lastpage}}

\maketitle{}

\begin{abstract}
Active galactic nuclei (AGN) in the early Universe are thought to be prominent sources of energy and ionizing photons that affected the growth of their host galaxy and their environment. However, it is still unclear how the supermassive black holes (SMBHs) that fuel these AGN grew to the observed high masses already at high redshifts. Observations of high-redshift SMBH progenitors or lower-luminosity AGN will thus help characterize the evolution of SMBHs and their impact on the surroundings. With the launch of the \textit{JWST}, fainter objects at high redshifts can now be detected, including lower-luminosity AGN. We assess the observability of such low luminosity AGN, using the cosmological simulation code \textsc{gizmo} to provide a realistic environment for black hole growth in the early Universe. Soon after the first stars are born in the simulation run, we insert stellar-remnant black hole seeds of various initial masses, between $300$ and $10^4 {\rm \ M}_{\odot}$, at the center of a dark matter halo and follow their growth until $z\sim6$. Such stellar black hole seeds placed in a typical high-$z$ environment do not significantly accrete and grow to reach masses that can be observed with the \textit{JWST} under conditions of standard Bondi-Hoyle accretion, as energy input from stellar feedback and chaotic dynamics prevent efficient gas accretion onto the black holes. To be observed with the \textit{JWST}, rarer but still physically feasible growth regimes, involving Eddington or super-Eddington accretion, would be required. Alternatively, AGN observability may be boosted under even rarer conditions of extreme gravitational lensing.
\end{abstract}





\section{Introduction}\label{1}
Observations of active galactic nuclei (AGN) at high redshifts have been pursued to answer questions such as how the first supermassive black holes (SMBHs) grew to massive sizes around $\sim10^9$ M$_\odot$ already so early in cosmic history at $z\gtrsim 6$ \citep[e.g.][]{Wu2015,Banados2018,Zubovas2021,Fan2022}. Under the standard black hole formation and growth channels, starting from a stellar remnant with growth limited by the theoretical maximum of the Eddington rate for radiatively efficient accretion, SMBHs at high redshifts would not have had enough time to grow as massive as required by observations \citep{Smith2019_2,Inayoshi2020}. Thus, understanding SMBH formation in the early Universe remains a fundamental challenge \citep[][]{Volonteri2021}. 

Furthermore, observations show various relationships between SMBHs and their host galaxies: Prominently among them are the correlations between the SMBH mass and the galaxy bulge velocity dispersion \citep{Gebhardt2000,Graham2011,Kormendy2013}, luminosity \citep{Beifiori2012}, and mass \citep{Croton2006,Ding2020}. In this regard, simulations indicate that AGN feedback can affect star formation in galaxies, resulting in quenching when their interstellar medium (ISM) is heated, or enhancing it by pressurizing the ISM gas \citep[e.g.][]{Wagner2016,Shirakata2019,Valentini2021}. Upcoming high-redshift galaxy and AGN observations \citep[e.g.][]{Finkelstein2022,Curtis2022,Harikane2022,Donnan2022} will test whether these relationships persist at high redshifts, or are subject to change, thus serving as constraints that theoretical models and simulations will need to reproduce.

Finally, it is an open question to assess the role of AGN during reionization, in particular its later stages. Some studies suggest that AGN could have been a significant contributor of ionizing photons, comparable to stellar sources \citep[e.g.][]{Madau2015,Volonteri2016,Jeon2022}. However, the general view is that the AGN number density is too low at high redshifts to complete reionization, and that galactic UV sources dominate \citep[e.g.][]{Ciardi2003,Robertson2015,Parsa2018,Finkelstein2019,Naidu2020}. Again, upcoming high-redshift AGN observations, comprising a multi-waveband view, are needed to constrain AGN number densities and luminosities to better understand their contribution to reionization. 

Predictions had been made before the launch of \textit{JWST} that it could achieve numerous detections of high-redshift AGN \citep{Volonteri2017}, and the initial results from \textit{JWST} have provided such AGN candidates at $z\gtrsim 6$ \citep[e.g.][]{Ding2022,Larson2023,Onoue2023}, including spectroscopic confirmations \citep[][]{Eilers2023,Kocevski2022,Kocevski2023}, as well as greatly improved observations more locally \citep[e.g.][]{Alvarez2022,Lai2022}. The expectation is that deeper observations in the near future will reveal AGN at even higher redshifts. Following up on recent studies to guide the incoming \textit{JWST} observations of AGN/SMBH activity in the first billion years \citep[e.g.][]{Gilli2022,Goulding2022,Lyu2022,Oogi2022,Windhorst2022}, we here focus on the lower luminosity AGN expected in the more typical, and thus lower-mass, host haloes at high redshift. Any constraints on this hitherto undetected population can provide key insights on the formation and evolution of the first galaxies \citep{Bromm2011}.

The current understanding of early SMBH formation proposes two channels for seeding it \citep[e.g.][]{Haemmerle2020,Sassano2021}. The first relies on seed black holes from the remnants of metal-free stars, the so-called Population~III (Pop~III), which are predicted to favor massive stars \citep[e.g.][]{Stacy2016,Hirano2017,Latif2022}. As a consequence of the top-heavy Pop~III initial mass function (IMF), the remnant black holes may thus also be quite massive, with $\sim10^2-10^3$ M$_\odot$, subsequently growing further through accretion and possibly mergers \citep{Heger2003}. A second, less common pathway, invokes more massive seeds, of order $\sim10^5$ M$_\odot$, the so-called direct collapse black holes (DCBHs), reflecting their origin in the collapse of massive primordial gas clouds, under rare conditions that allow the gas to collapse without forming stars \citep[e.g.][]{Bromm2003,Wise2019}. Observations of these seed black holes at high redshift, as they are believed to be the progenitors of the massive $\sim10^9$~M$_\odot$ SMBHs that we observe less than a billion years after the Big Bang, will provide key insights to the questions outlined above. Previous work suggests that the massive end of the early AGN population can be detected, if unobscured by their environment \citep{Gilli2022,Goulding2022}, but the observability of the more common stellar mass black holes and lower-luminosity AGN at high redshifts ($z\sim6-8$) is yet to be tested extensively.

This paper is organized as follows: In Section~\ref{2}, we briefly describe the simulation code used, together with the modifications made to the previous setup in \citet{Liu2020}, including our new recipe of inserting central SMBH seeds in the simulation box soon after the first stars formed. In Section~\ref{3}, we analyze our simulations in terms of the resulting growth of the seed masses. In Section~\ref{4}, we discuss the corresponding observability with \textit{JWST}, for the suite of SMBH seeds considered here. We summarize our findings and offer conclusions in Section~\ref{5}.
\\


\section{Numerical Methodology}\label{2}

We assess the observability of the lower mass, low-luminosity AGN by incorporating seed SMBH growth in a cosmological environment using numerical simulations, following up on earlier hydrodynamic and N-body simulations of AGN growth and their luminosity functions \citep[e.g.][]{Somerville2015,Griffin2019,Oogi2022}. We specifically use the \textsc{gizmo} code \citep{Hopkins2015}, which combines new Lagrangian solvers for hydrodynamics with the gravity solver inherited from the \textsc{gadget-2} code \citep{Springel2005}. We use the modified version of \textsc{gizmo} developed by \citet{Liu2020} with the Lagrangian meshless finite mass (MFM) hydro solver. This version of the code implements updated sub-grid prescriptions for star formation, stellar feedback, black hole formation, accretion, and feedback on top of the models for primordial chemistry, cooling, and metal enrichment in \citet{Jaacks2018,Jaacks2019}. Unlike \citet{Liu2020}, rather than focusing on the overall properties of the simulation box, we focus on the most massive host halo in a zoom-in region, placing black hole seeds of various masses in its center after the first stars have already formed in the simulation run. With the established routines for black hole accretion \citep{Liu2020}, we follow the growth of the central seed black holes to predict their fluxes that will reach us today.

\subsection{Initial simulation setup}

We specifically employ the initial parameters and simulation setup of the \texttt{FDzoom\_Hseed} run in \citet{Liu2020}, which targeted a zoom-in region defined around a halo with (total) mass $\sim10^{10}$ M$_\odot$ at $z \sim 8.5$, with a comoving volume $V_C \sim4\ h^{-3}$ Mpc$^3$ \citep{Liu2019}. Initial conditions are generated with the \textsc{MUSIC} code \citep{Hahn2011} at a redshift of $z=99$, using \textit{Planck} cosmological parameters \citep{Planck2016}: $\Omega_m = 0.315$, $\Omega_b = 0.048$, $\sigma_8 = 0.829$, $n_s = 0.966$, and $h = 0.6774$. The initial mass of a gas particle (without losing mass to form stars) in the simulation is $m_{\rm gas}\simeq 6400\ h^{-1}\rm M_{\odot}$ and its (comoving) gravitational softening length $0.2\ h^{-1}\rm kpc$. The physical spatial resolution at $z\sim 10$ is thus $\sim$30\,pc. The full details of the star formation, black hole physics, and other aspects of the simulation run can be found in \citet{Liu2020}, which we closely follow here. We will describe below how we deviate from the previous work in order to test the observability of AGN in the first galaxies.

To study the growth of stellar seed black holes, we run the simulation until $z=15$, soon after the first star particles form in the simulation box. We then stop the simulation, identify haloes with \textsc{Rockstar} \citep{Behroozi2013} in post-processing, and select the most massive one at this time as our target. We note that this halo is distinct from the one that grows into the $10^{10}$ M$_\odot$ halo at $z\sim8.5$ that was used to define the zoom-in region. We further point out that halo mass estimates may be affected by the use of different methods to identify the dark matter halos, as the parent simulation employed the \textsc{Caesar}\footnote{http://caesar.readthedocs.io/en/latest/} code instead of \textsc{Rockstar}. We further identify the densest gas particle contained within the target halo viral radius with \textsc{yt} \citep{Turk2011}, choosing that particle to represent the central seed black hole by manually converting it to a black hole sink particle, following the routine of \citet{Liu2020}. We explore three cases of different stellar seed black hole masses. The first case is based on the halo-central black hole mass relation from \citet{Jeon2022}, derived by assuming an energy balance between the gravitational potential of the halo and the radiation feedback from the black hole, resulting in a seed black hole mass of $\sim3\times10^2$ M$_\odot$. The other two consider more massive seeds, with $10^3$ M$_\odot$ and $10^4$ M$_\odot$. The latter case can be regarded as the extreme upper limit for black holes of stellar origins, invoking collision scenarios in primordial star clusters \citep[e.g.][]{Katz2015,Reinoso2023}, which is close to the typical mass of a gas particle in the simulation. While this implies that mass conservation is not strictly enforced for our first two seeding cases, we only change one particle mass in this process, and such stellar seed black hole masses are much smaller than the typical gas particle mass in the simulation box, so that this effect is negligible.

\subsection{Black hole accretion}
We model the most optimistic growth of the seed black holes to test the upper limit of observability. To do so, we use the original Bondi-Hoyle equation to determine black hole accretion \citep{Bondi1944}, without considering rotation or gas velocity near the black hole:
\begin{equation}
\dot{M}_{\rm acc} = \frac{4\pi(GM_{\rm BH})^2\rho_g}{c_s^3}   \mbox{\ ,}
\label{bondi}
\end{equation}
where $M_{\rm BH}$ is the black hole mass, $\rho_g$ the gas density within the hydro kernel at the black hole position, and $c_s$ the sound speed. Given this value of $\dot{M}_{\rm acc}$, we add at each simulation time step $\delta M =\dot{M}_{\rm acc}\delta t $ to the black hole mass, where $\delta t$ is the size of the local time step. The dynamical mass of the black holes, which represents the total sink particle mass, in general including neighboring stars, their remnants, and any surviving gas, bound to the black hole, are smoothly updated as well. For simplicity, we here set the dynamical mass of the sink particle equal to the black hole mass. Furthermore, while \citet{Liu2020} require black holes with masses greater than $10^4$ M$_\odot$ to only update their dynamical masses when they stochastically swallowed gas particles, we update their dynamical masses even when they do not swallow a gas particle. This is assuming that the black hole can accrete mass continuously so that the most optimistic accretion can be followed.

Again, as the total number of black holes formed is small, including the one we explicitly insert, and the black hole masses are typically much smaller than the gas particle masses, any violation of mass conservation due to the increased black hole masses is not significant.

\subsection{Feedback physics}
We include the stellar and thermal black hole feedback from \citet{Liu2020}. The simulation cannot resolve individual stars, so that stellar particles represent a stellar population, characterized by the IMF for Pop~III and Pop~II from \citet{Jaacks2018,Jaacks2019}. A global Lyman-Werner (LW) background from the Pop III/II stars is calculated, based on the star formation rate density at a specific time. Furthermore, a local LW field is applied in the neighborhood of a newly born star for a given period. The total LW radiation at time $t$ and position $\vec{x}$ is thus the sum of the global and local fields:
\begin{equation}
    J_{\rm LW}(t,\vec{x}) =  J_{\rm LW,global}(t)+ J_{\rm LW,local}(t,\vec{x})\mbox{\ .}
\end{equation}

Following \citet{Jaacks2019}, we include a global photoionization heating effect, which arises from a redshift-dependent photoionization rate and is caused by the UV background produced by stars \citep{Faucher2009}. Local photoionization heating is also applied to the gas particles within the Str\"{o}mgren radius of active star particles on-the-fly. Similar to the LW feedback, the total photoionization heating is calculated as the sum of the global and local components. 

For supernova (SN) feedback, the simulation resolution cannot resolve individual SN explosions. Therefore, when a star particle dies after a typical stellar lifetime (3~Myr for Pop~III and 10~Myr for Pop~II), the total produced metals are distributed evenly around the final radius of the expanded SN shell. This radius depends on the total energy of SN explosions in the stellar population. For Pop~III, the SN energy and mass of produced heavy chemical elements are calculated on-the-fly by counting progenitors sampled from the IMF - core-collapse and pair-instability SNe. For Pop II stars, IMF integrations are used to calculate the SN energy as $E_{\rm SN} \simeq 10^{52}$ erg $\times~m_\star/(10^3$ M$_\odot)$ and the produced metals as $M_Z = 0.016m_\star$, where $m_\star$ is the mass of the stellar population represented by the particle. In addition, we also model the thermal feedback for Pop~III SN explosions by injecting thermal energy - increasing the temperature by $2\times10^4$~K - and instantly ionizing the hydrogen of the gas particles within the Str\"{o}mgren radius, after a Pop~III stellar particle reaches the end of its lifetime.

We also include SN-driven winds from Pop~II stars to account for mechanical feedback \citep{Springel2003}. The probability for a Pop~II candidate gas particle to be launched as a wind just before spawning a stellar population is calculated as
\begin{equation}
    p_w = 1-\exp\left(-\eta_{w,{\rm SF}}\frac{m_\star}{m_{\rm SF}} \right) \mbox{\ ,}
\end{equation}
where $m_{\rm SF}$ is the gas particle mass and $\eta_{w,{\rm SF}}=2$ the wind-loading factor. A random number is generated to sample the probability of launching. If launched as a wind, the gas particle receives a kick of $\simeq240$ km s$^{-1}$ in a random direction \citep{Springel2003}. The ejected gas particle is decoupled from hydrodynamics and cannot form stars, before a time of $0.1H(z)^{-1}$ has passed or its (hydrogen number) density has dropped below $10\ \rm cm^{-3}$, where $H(z)^{-1}$ is the local Hubble time. This model is not applied for Pop~III stars, as the full impact of SN feedback has already been captured by the photoionization heating and thermal energy injection.

Finally, we include thermal and mechanical feedback from black holes based on \citet{SMH2005,Tremmel2017,Negri2017}. Thermal feedback is implemented by energy injection into the gas particles near the black hole. The total energy to be distributed to the gas particles for a given time step $\delta t$ is $\delta E = \epsilon_r L_{\rm BH}\delta t$, where $\epsilon_r=0.02$ \citep{Tremmel2017} is the radiation–thermal coupling efficiency and $L_{\rm BH}$ the black hole luminosity calculated via 
\begin{equation}
    L_{\rm BH} = \epsilon_{\rm EM}\dot{M}_{\rm acc}c^2\mbox{\ .}
\end{equation}
Here, the radiative efficiency $\epsilon_{\rm EM}$ is defined as
\begin{equation}
    \epsilon_{\rm EM} = \frac{\epsilon_0A\eta}{1+A\eta},~\eta\equiv\dot{M}_{\rm acc}/\dot{M}_{\rm Edd}\mbox{\ ,}
\end{equation}
following \citet{Negri2017}, with $\epsilon_0=0.125$ and $A=100$. The Eddington accretion rate is evaluated as 
\begin{equation}
\dot{M}_{\rm Edd} = 2.7\times10^{-5}\left(\frac{M_{\rm BH}}{1000~\text{M}_\odot}\right)\left(\frac{\epsilon_0}{0.1}\right)^{-1}\rm~M_\odot~\text{yr}^{-1}\mbox{\ .}
    \label{edd}
\end{equation}
The mechanical feedback only applies to black holes with masses greater than $10^4$ M$_\odot$, when the black hole can swallow nearby gas particles stochastically. A wind gas particle is produced following the prescription in \citet{Negri2017} when a gas particle is swallowed, where the wind is launched with a kick velocity opposite to the in-fall direction of the swallowed particle, given by 
\begin{equation}
    v_w = 200~\text{km s}^{-1}\left(\frac{M_{\rm BH}}{10^4~\text{M}_\odot}\right)^{1/2}\mbox{\ .}
\end{equation}

The feedback modules described above affect the temperature and density of the gas particles near the black hole, thus impacting its accretion rate and growth. The full details and equations for the feedback physics can be found in \citet{Liu2020}. It will be shown below that since black holes in our simulations cannot grow significantly above $10^{4}\ \rm M_{\odot}$, mechanical feedback is unimportant.

\section{Results}\label{3}

From the simulation output, we track the growth of the target black holes inserted into the run at $z=15$. We focus on the evolution of the mass and the accretion rate, comparing them to the theoretical maximum of their respective Eddington rates. In the following, we consider the two classes of black hole seeds, light (stellar) and heavy (direct-collapse).

\subsection{Black hole growth}

\begin{figure*}
    \centering
    \begin{multicols}{3}
    \includegraphics[width=0.33\textwidth]{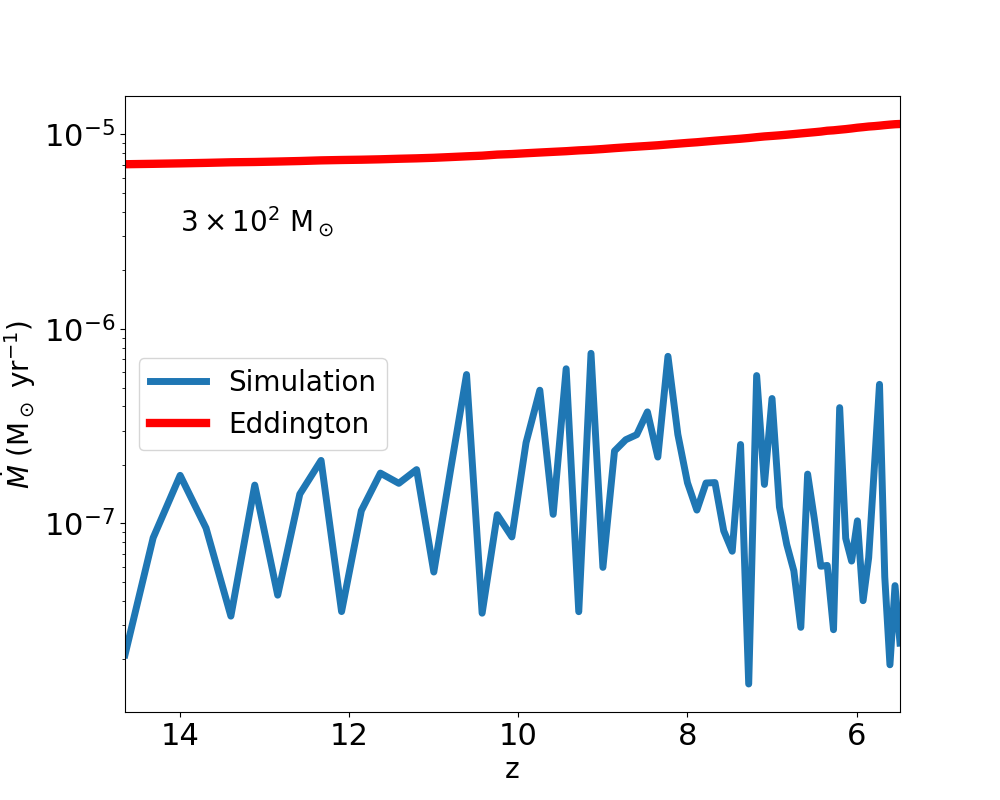}\par
    \includegraphics[width=0.33\textwidth]{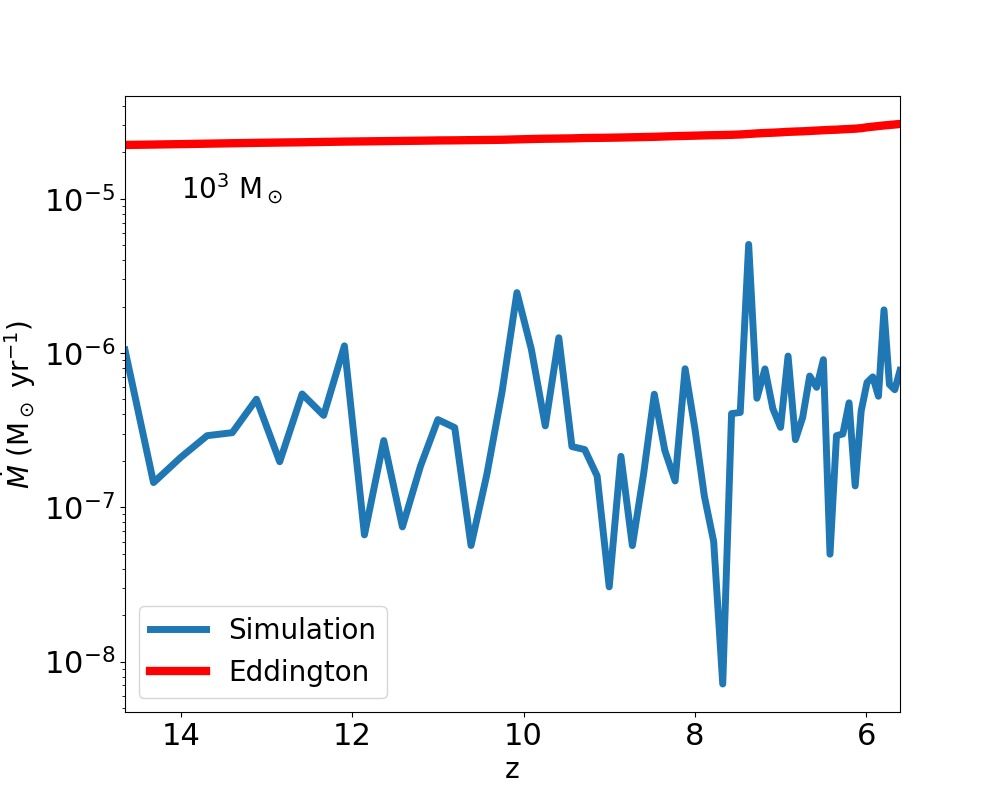}\par
    \includegraphics[width=0.33\textwidth]{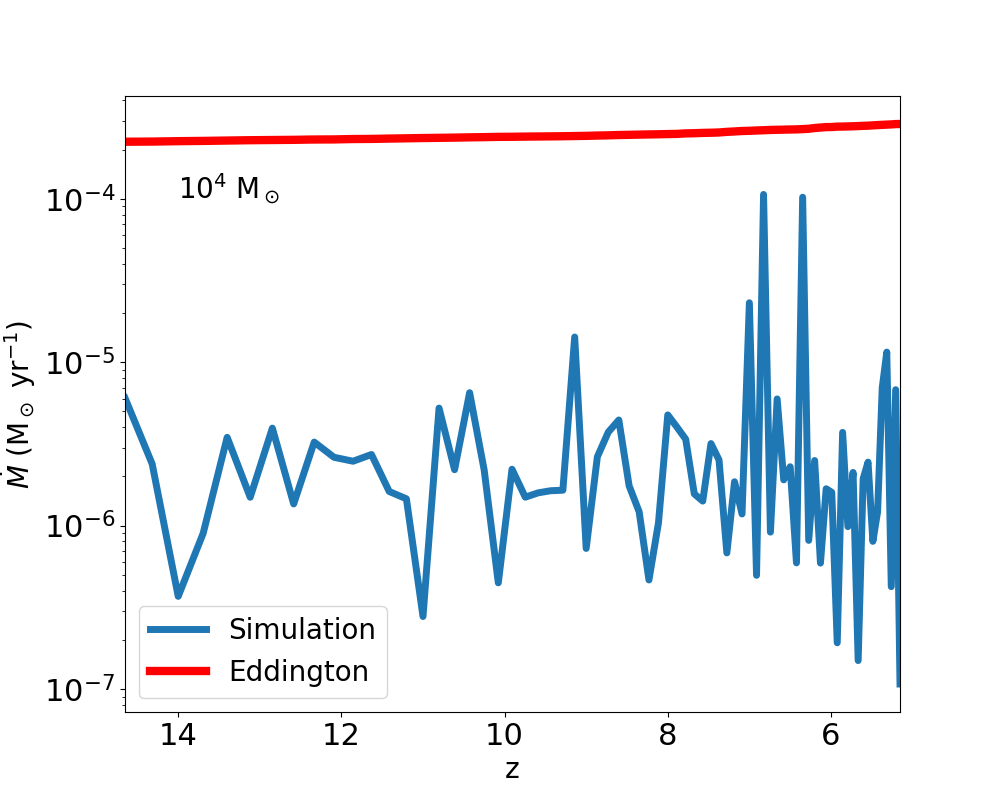}\par
    \end{multicols}
     \begin{multicols}{3}
    \includegraphics[width=0.33\textwidth]{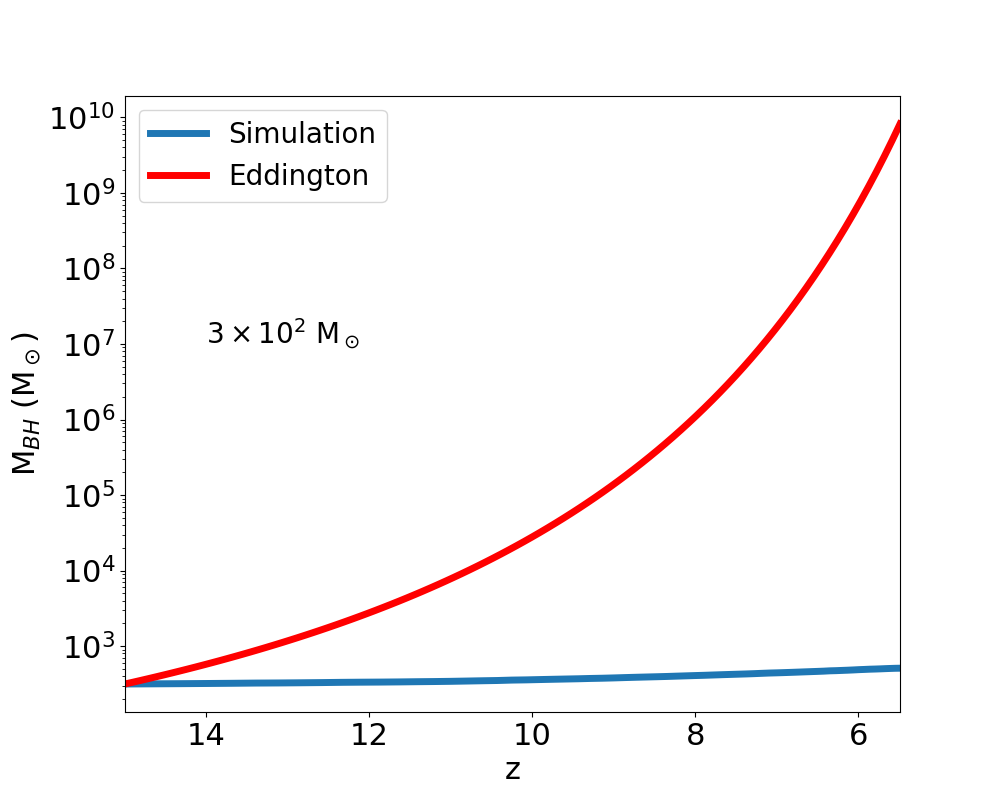}\par
    \includegraphics[width=0.33\textwidth]{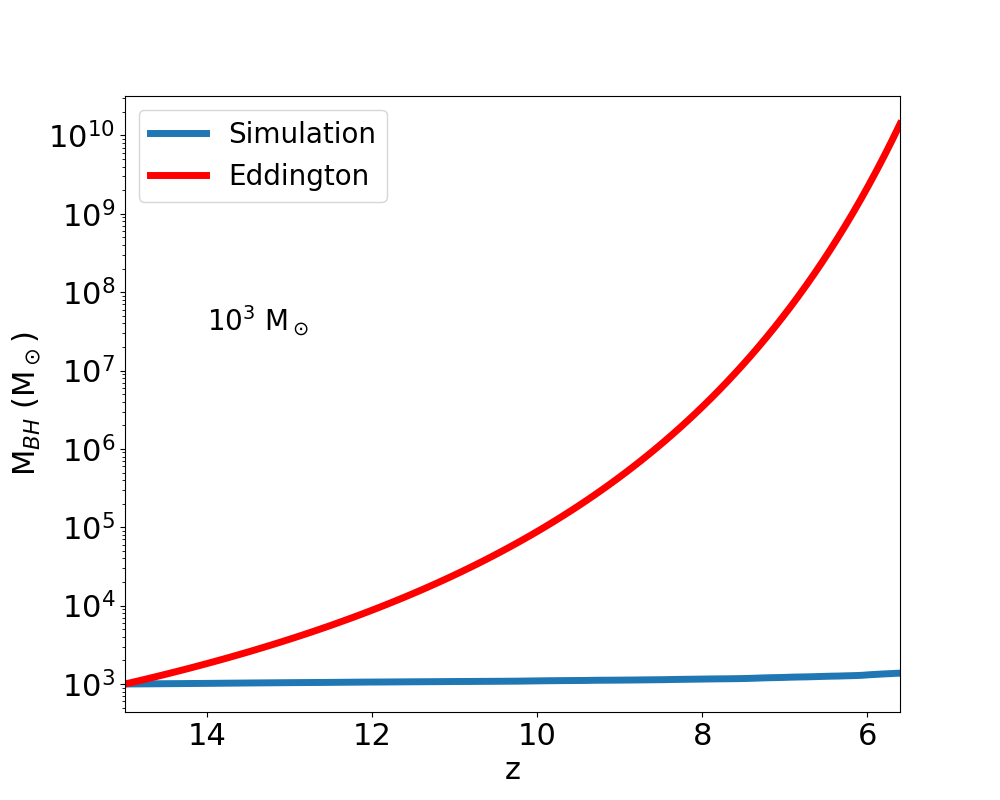}\par
    \includegraphics[width=0.33\textwidth]{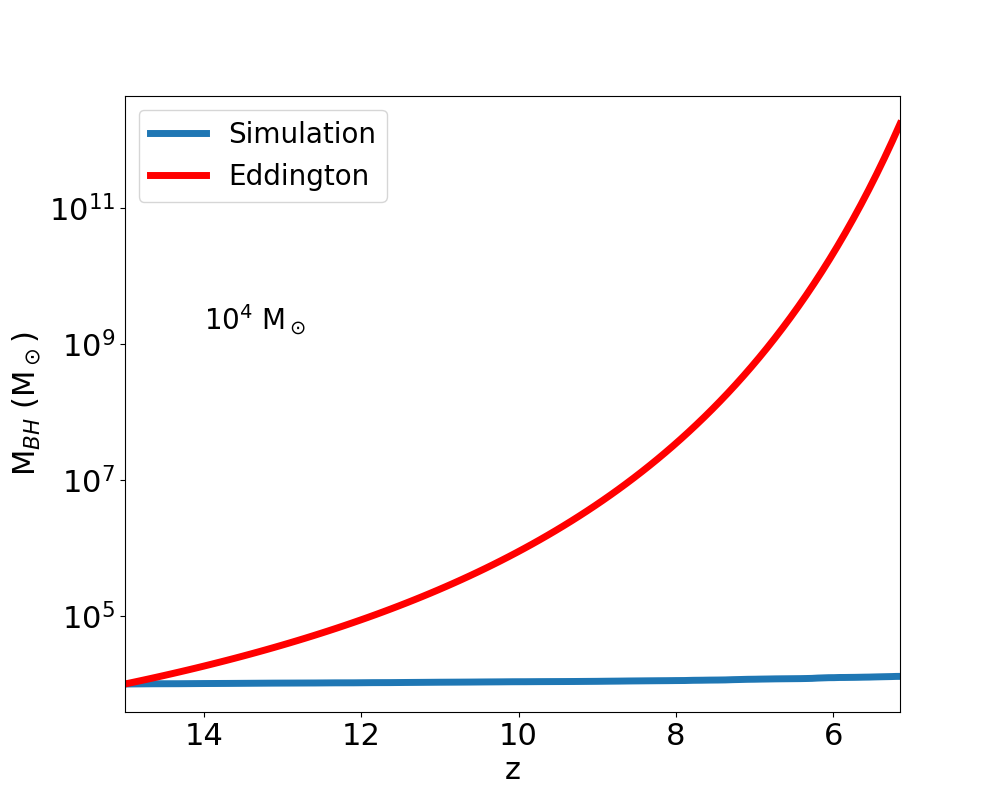}\par
    \end{multicols}
    \caption{Growth of stellar seed black holes. {\it Top panels:} Mass accretion rate vs. redshift. {\it Bottom panels:} Black hole mass vs. redshift. We consider seed masses of $\sim3\times10^2$ M$_\odot$ (left), $10^3$ M$_\odot$ (middle), and $10^4$ M$_\odot$ (right). The mass accretion rate inferred from the \textsc{gizmo} simulations (blue) is compared to the Eddington rate, evaluated for the masses reached in the simulations at each redshift (red). The black hole masses encountered in the simulations (blue) are plotted together with the masses, achieved if growth had proceeded at the Eddington rate from the start (red). Evidently, accretion rates for all seed masses remain significantly below Eddington for all redshifts, and on average stay constant. As a consequence of the sub-Eddington conditions, seed masses remain close to their initial values (see Fig.~\ref{stellarseedratio}).}
    \label{stellarseed}
\end{figure*}

\begin{figure*}
    \centering
    \begin{multicols}{2}
    \includegraphics[width=0.5\textwidth]{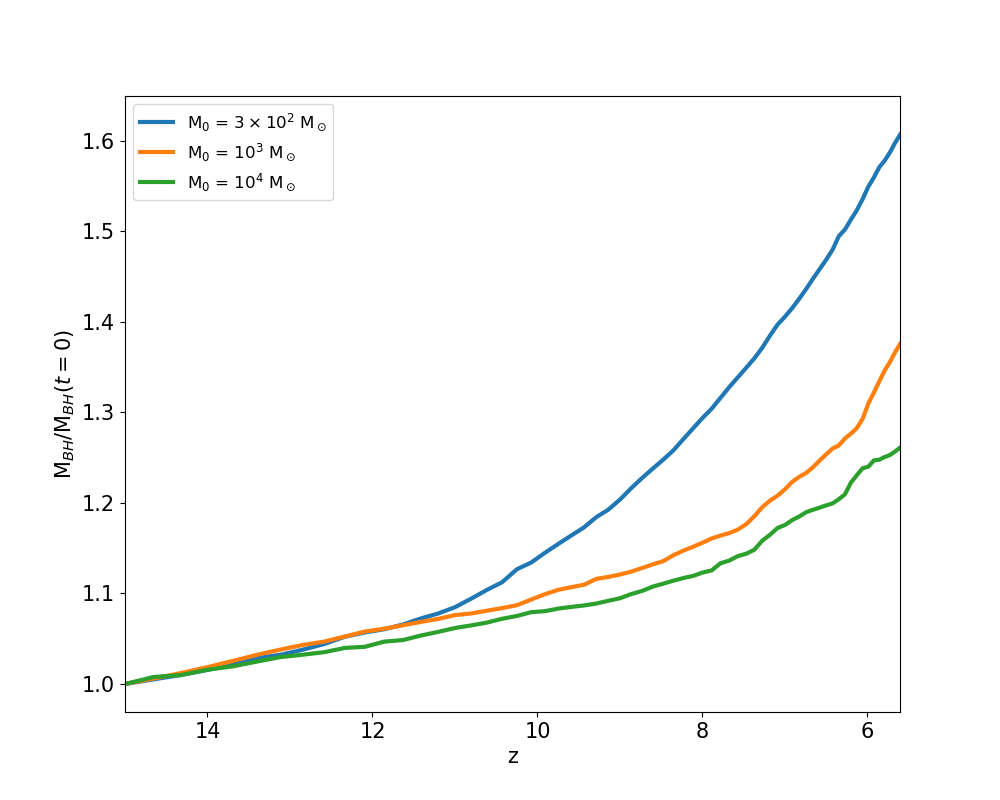}\par
    \includegraphics[width=0.5\textwidth]{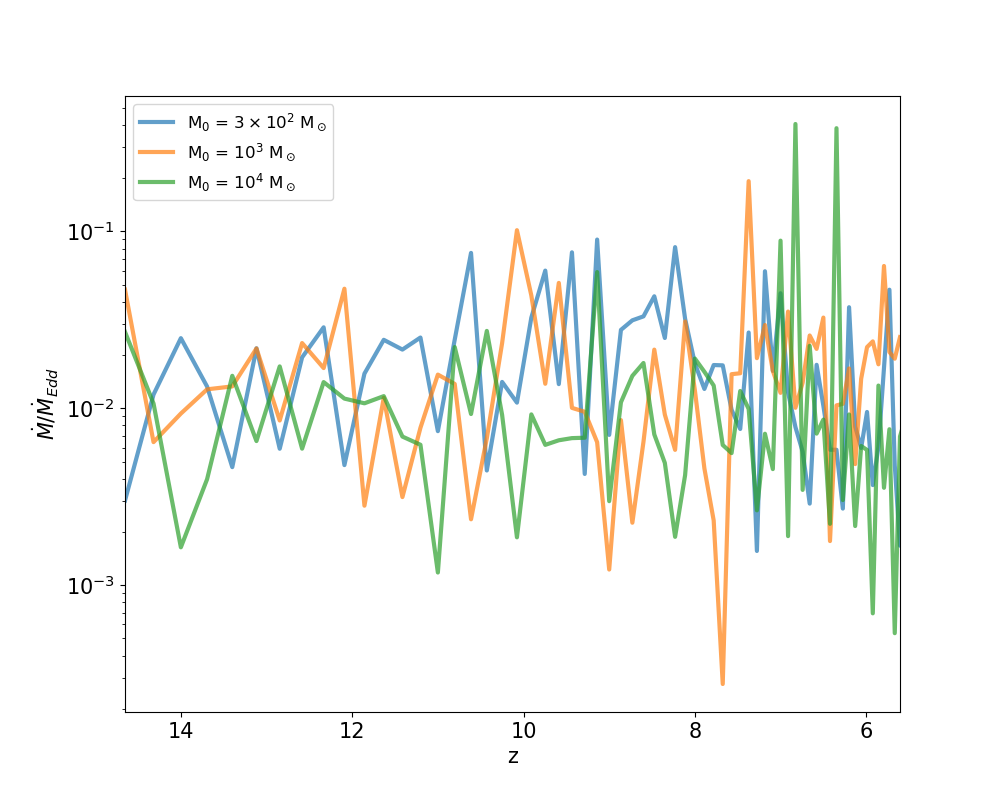}\par
    \end{multicols}
    \caption{Summary of stellar seed growth. {\it Left:} Ratio of black hole mass to the initial value vs. redshift, for the three cases considered. {\it Right:} Corresponding ratio of the mass accretion rate to the Eddington rate vs. redshift. As can be seen, all simulated growth trajectories are highly stunted, with strong variability in the Eddington ratios (right). Note the near-universal average ratios, across all seed masses, of around $\sim10^{-2}$ throughout the duration of the simulation.}
    \label{stellarseedratio}
\end{figure*}

We show the accretion rates and the growth of the three target stellar seed black hole masses in Figure~\ref{stellarseed}. The accretion rates for all the three cases remain below the Eddington rate, given in Equ.~\ref{edd}, for all redshifts. Consequently, their masses do not grow significantly, staying within a magnitude of their initial values. To further analyze the seed black hole growth, we in Figure~\ref{stellarseedratio} consider the ratio of the increasing black hole mass to the initial value, as well as that of the mass accretion rate to the Eddington rate across redshifts.
As can be seen, the black holes accrete around 40\% of their initial masses by the end of the simulation at $z\sim5.6$, after $\sim 700$\ Myr of accretion, and their accretion rates tend to remain at $10^{-2}$ times the Eddington rate for all initial seed black holes. Consequently, stellar seeds do not experience any significant growth, even for such a prolonged duration of accretion activity. Below, in Section~\ref{stellarfeedback}, we will explore the physical reasons for such stunted growth.

\begin{figure}
    \centering
    \includegraphics[width=0.5\textwidth]{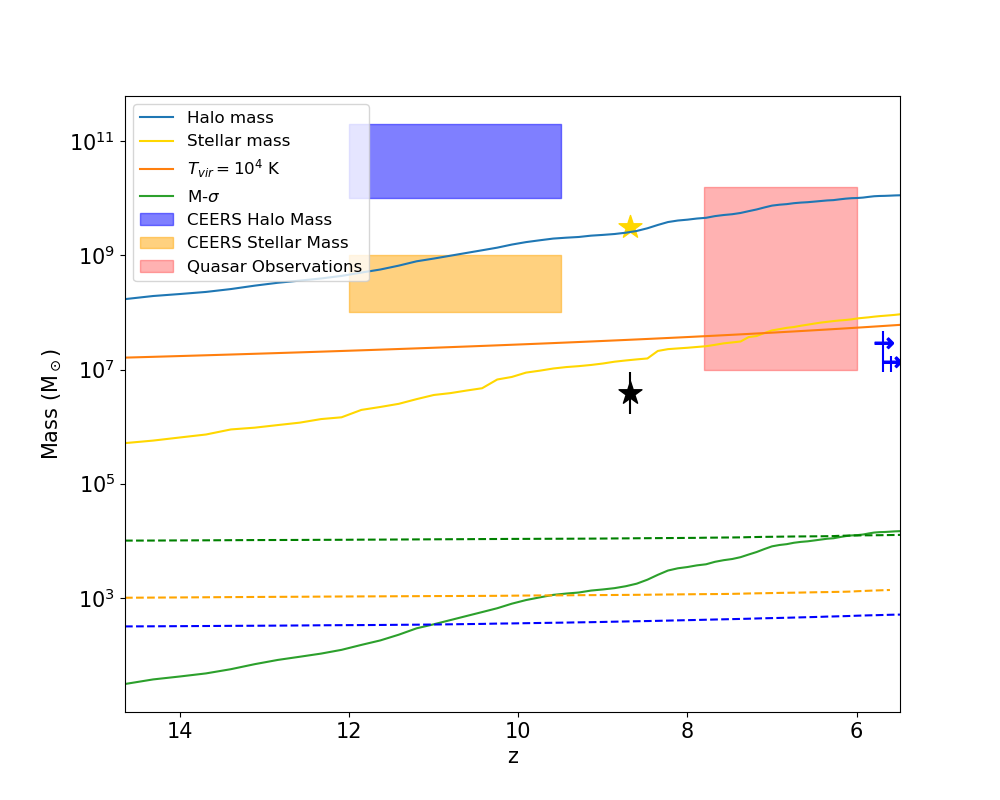}\par
    \caption{Mass growth trajectories for the first galaxies, hosting the black hole seed, vs. redshift. Specifically, we plot the virial (total) mass, as well as the stellar mass of the host halo. For reference, we reproduce the mass of an atomic cooling halo with a virial temperature of $10^4$\,K \citep{Bromm2013}, the halo and stellar masses observed in the \textit{JWST} Cosmic Evolution
Early Release Science (CEERS) survey \citep{Finkelstein2022}, as well as the expected SMBH mass for our simulated halo, assuming the empirical $M-\sigma$ relation \citep{Evrard2008,Kormendy2013}. Comparing with the $M-\sigma$ mass expectation, we show our simulated black hole mass growth (dashed lines in green, yellow, and blue, corresponding to $10^4, 10^3,$ and $3\times10^2$ M$_\odot$ seed masses, respectively). We further indicate the mass and redshift range of previously observed massive quasar black holes \citep[red region;][]{Inayoshi2020}, and mark two spectroscopically confirmed AGN observed with \textit{JWST} at $z\sim5.5$ \citep{Kocevski2023}. The \textit{JWST} observations are indicated by the blue arrows to the right as their actual redshifts are beyond the range of the plot. The black star shows the SMBH mass of the highest-redshift AGN currently observed, and the yellow star the stellar mass of the AGN host \citep{Larson2023}. For our simulated galaxy, stellar mass is a significant component, and thus it can be expected that stellar feedback is prevalent in the halo. We note that our simulated halo and black hole masses are lower than those of current observations, which are biased towards brighter and more extreme objects. In contrast, the halo and black hole masses studied here represent a more common case in the high-redshift Universe.}
    \label{halomass}
\end{figure}

\begin{figure*}
    \centering
 \begin{multicols}{2}
    \includegraphics[width=0.5\textwidth]{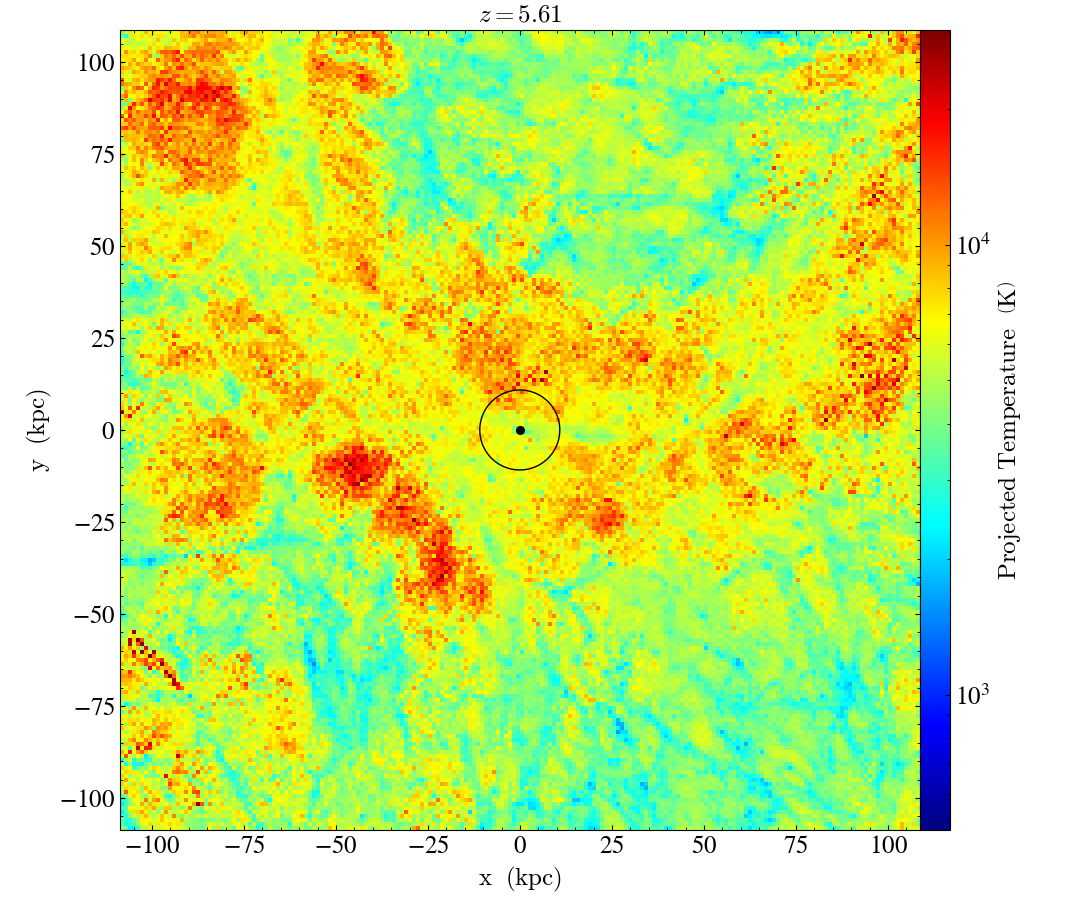}\par
    \includegraphics[width=0.5\textwidth]{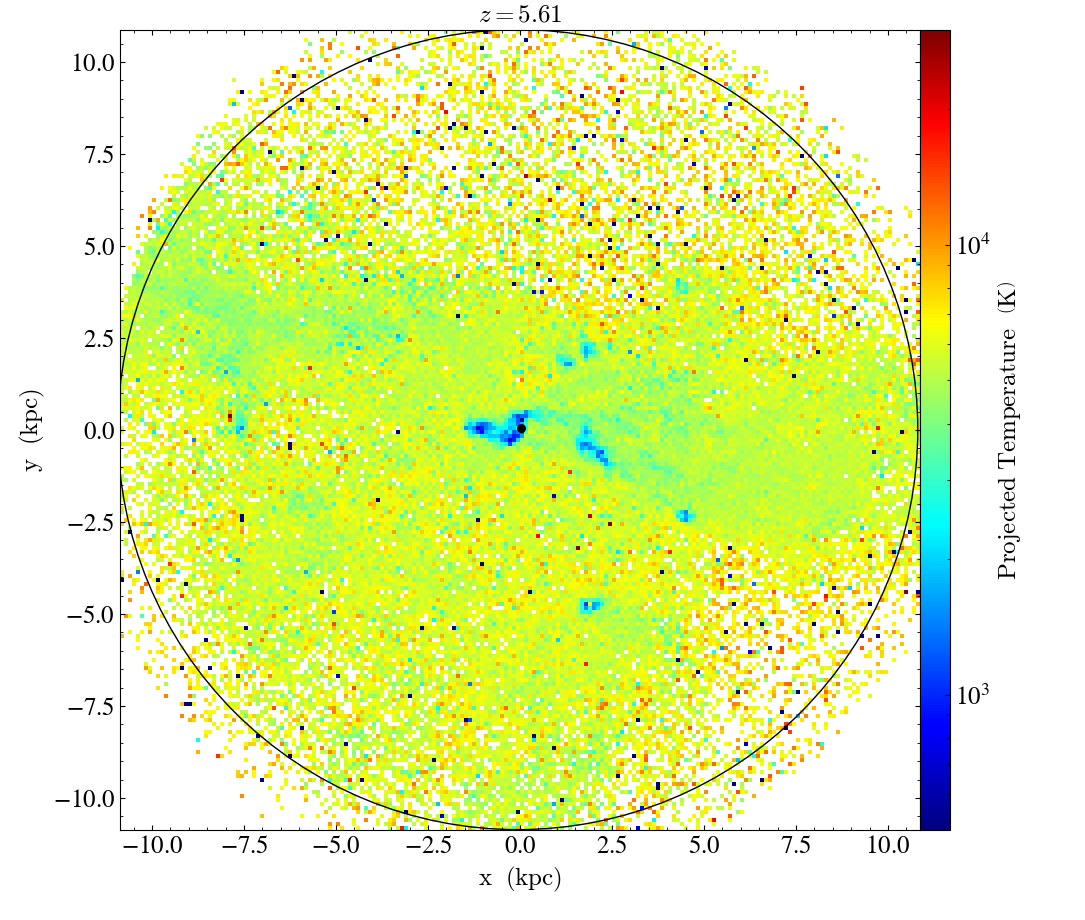}\par
    \end{multicols}
    \caption{Projected gas temperature in simulation run at $z=5.61$, employing physical coordinates. The values shown are the mass-averaged gas temperatures along the $z$-axis for the entire computational box (left) or only through the target halo virial diameter (right). The left panel shows the simulation region around the target halo hosting the black hole, and the right panel the target halo itself. The black circle marks the virial radius of the halo in which the target black hole resides. The black dot inside the circle marks the black hole position, with its size not to scale. The temperature in the halo and its environment is generally high as a consequence of the stellar feedback.}
    \label{temden}
\end{figure*}

\begin{figure*}
    \centering
    \begin{multicols}{2}
    \includegraphics[width=0.5\textwidth]{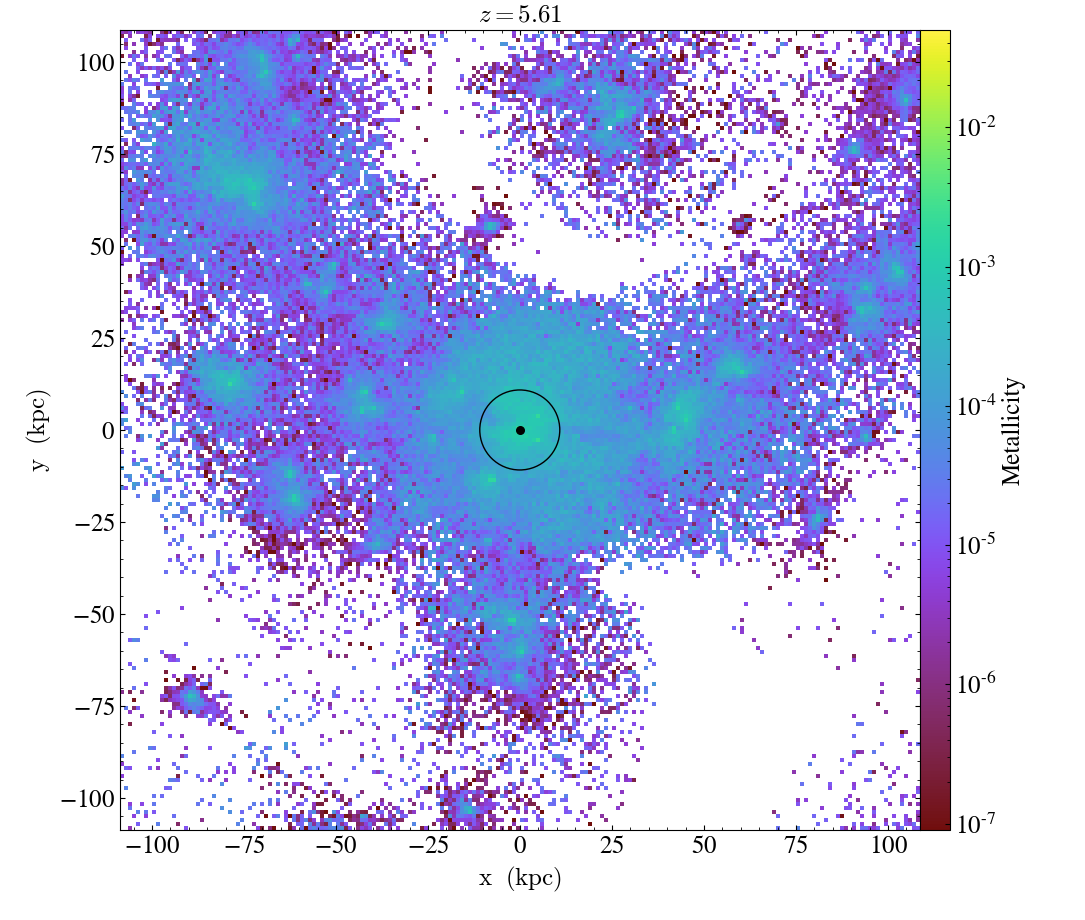}\par
    \includegraphics[width=0.5\textwidth]{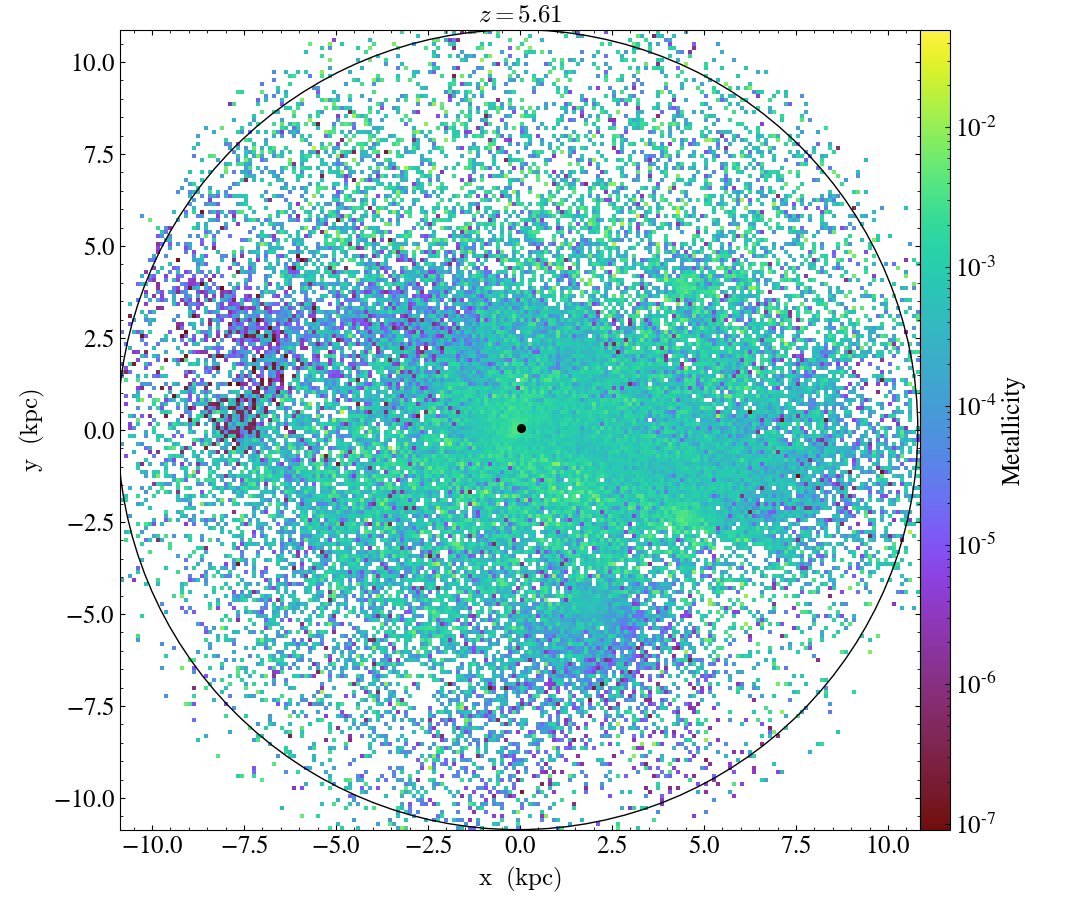}\par
    \end{multicols}
    \caption{Early metal enrichment from the first supernova explosions. The manner of presentation and the meaning of the black dot and circle are similar to Fig.~\ref{temden}, except that here (absolute) gas metallicities are shown, expressed as mass fractions. We again present mass-averaged quantities along the $z$-axis, as before. The metallicities are relatively high, reaching values close to solar, indicating significant prior star formation and correspondingly vigorous past supernova activity in the simulation box and halo.}
    \label{metal}
\end{figure*}

\begin{figure*}
    \centering
    \begin{multicols}{2}
    \includegraphics[width=0.5\textwidth]{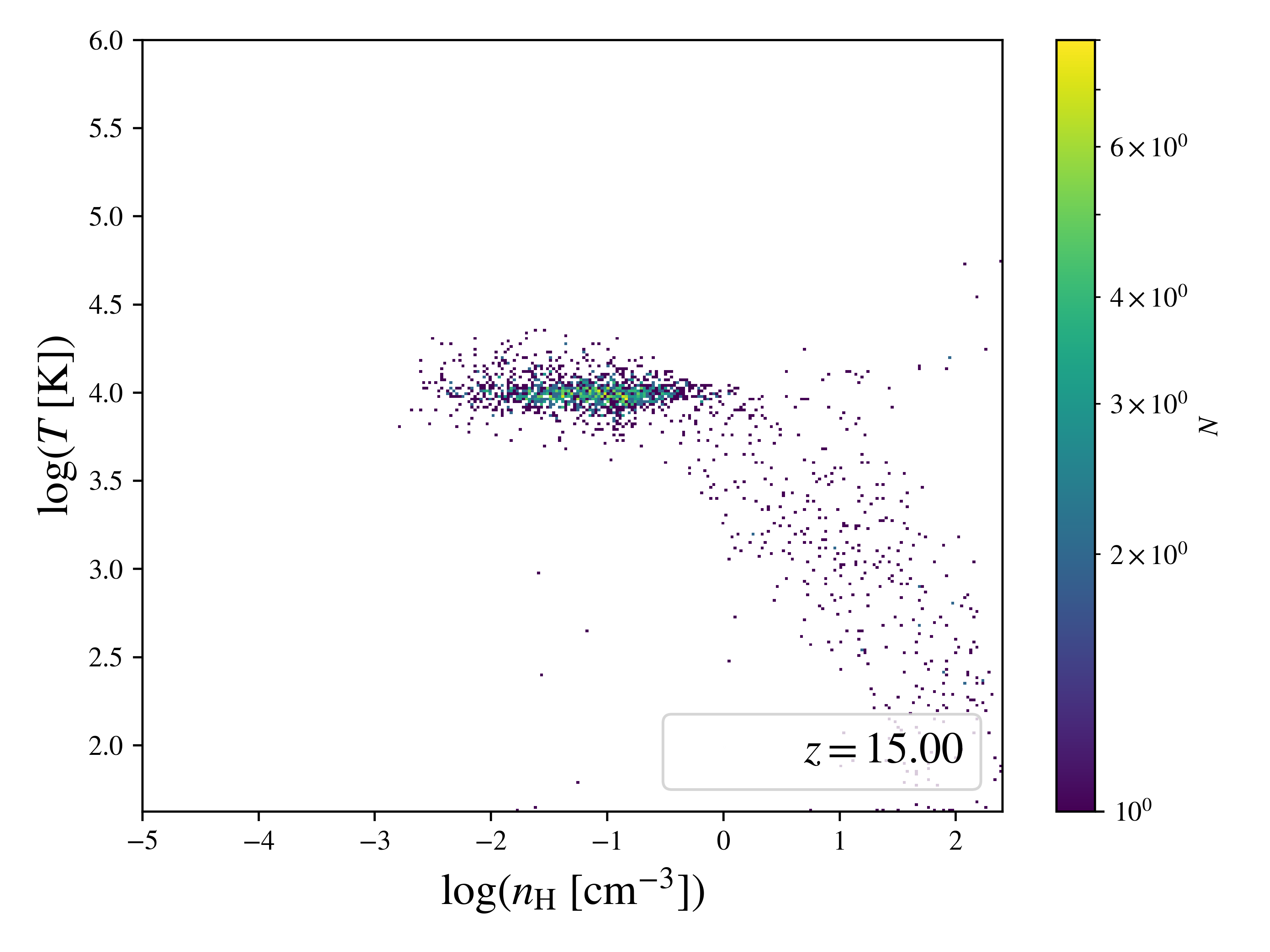}\par
    \includegraphics[width=0.5\textwidth]{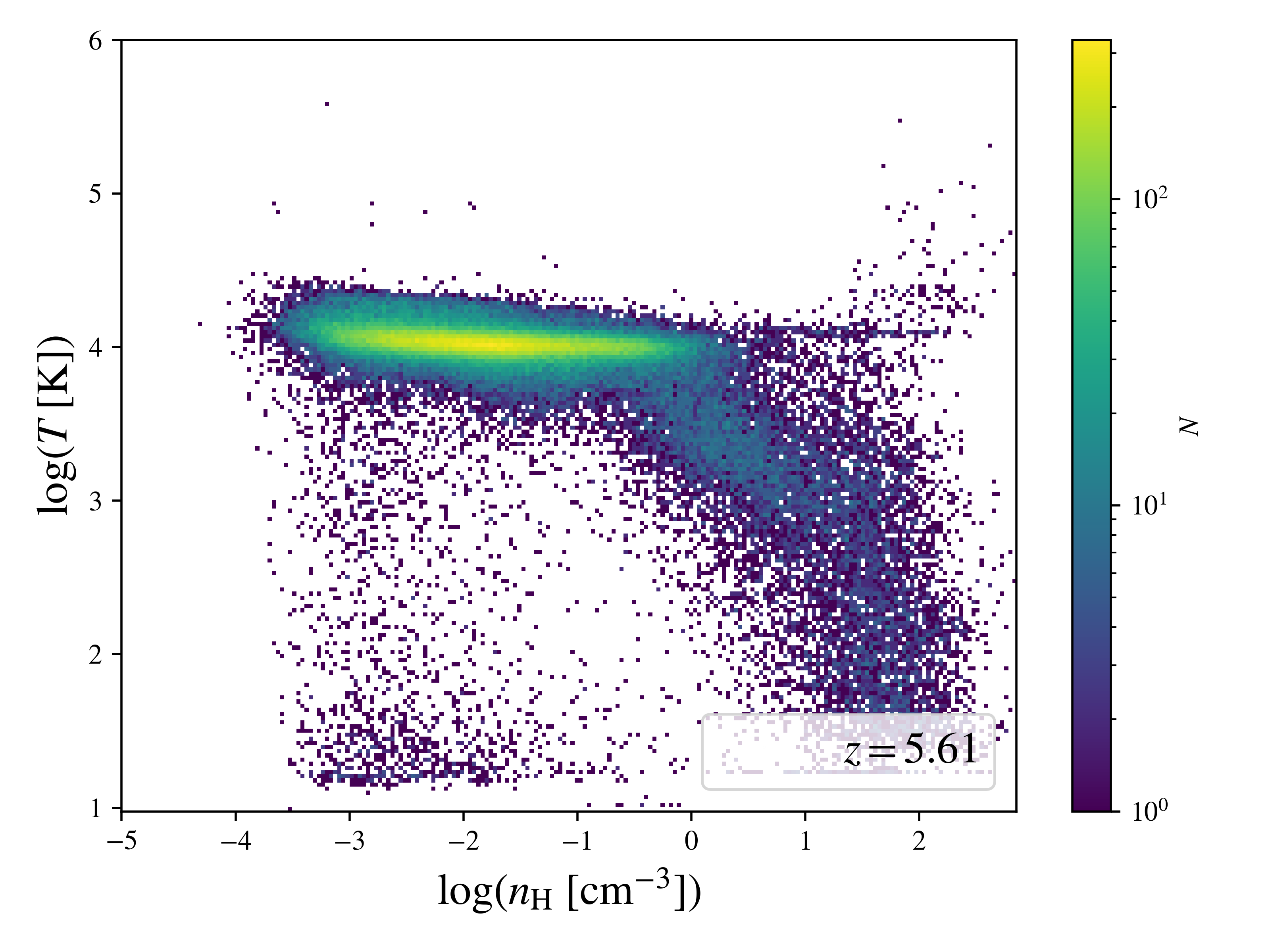}\par
    \end{multicols}
    \caption{Phase plot of the hydrogen gas temperature and (hydrogen) number density for the particles inside the target halo, as specified by the black circle in Fig.~\ref{temden}, at redshift 15 (left) and at 5.61 (right). The left panel shows the particle distribution just before the target black hole is inserted. As can be seen, particles with temperatures $\sim10^4$\,K are present, indicating that stellar feedback has already impacted the halo gas before $z\sim15$. At a later time, at $z=5.61$, the target halo contains mostly gas (photo-)heated by stars ($\sim10^4$\,K), together with some cold and dense star-forming gas ($\lesssim10^3$\,K). 
    Stellar feedback therefore is significant within the target halo, as suggested by the stellar mass estimate in Fig.~\ref{halomass}, acting to inhibit black hole growth.
    }
    \label{temphase}
\end{figure*}

\begin{figure}
    \centering

    \includegraphics[width=0.5\textwidth]{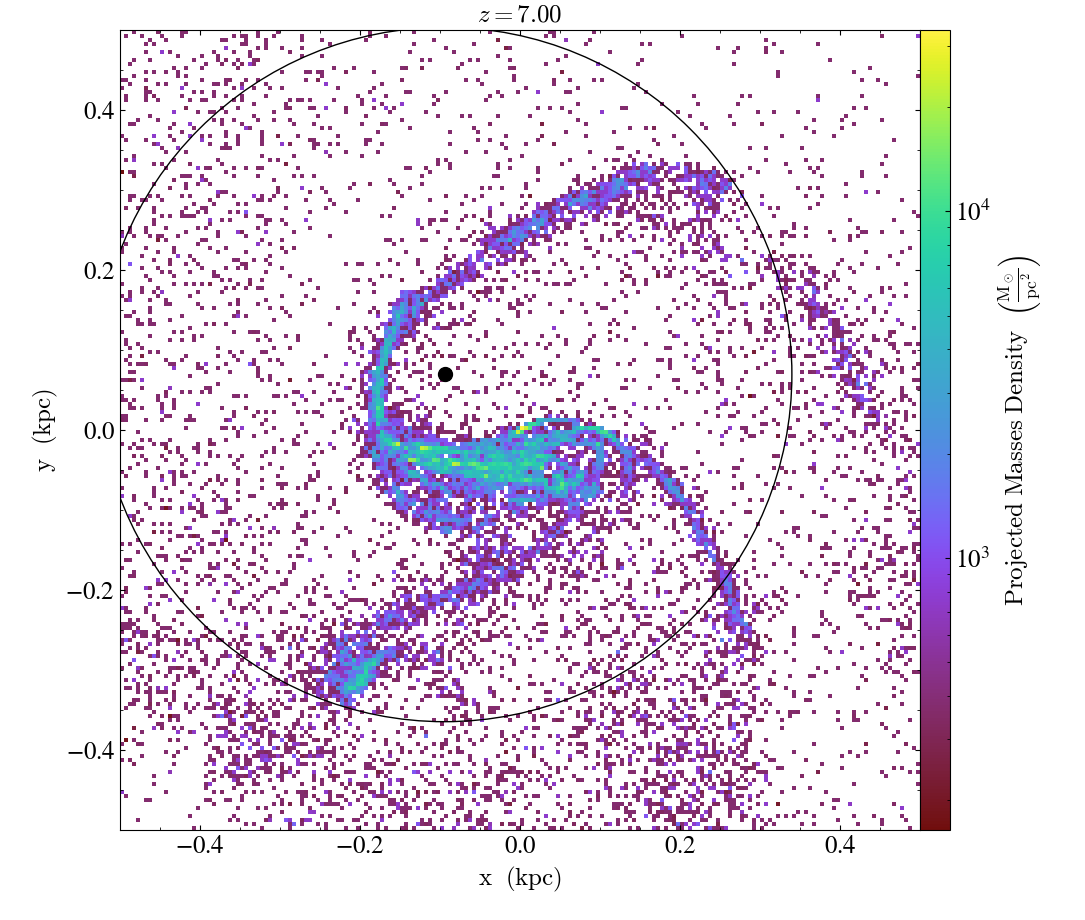}

    \caption{Gas surface density near the halo center at $z=7$ in the simulation run where stellar feedback and star formation are disabled in the vicinity of the black hole. The black circle marks the no-feedback region, and the black dot the black hole position, with its size not to scale. The black hole wanders away from the gas density peak at the center, likely due to perturbations from overly massive gas and dark matter particles under the limited resolution of our simulation. It is also possible that in reality, high-redshift low-mass galaxies do not have well-defined dynamical centers, such that a low-mass black hole cannot sink into a dense central region by dynamical friction \citep{Boldrini2020,Ma2021}.
    }
    \label{nofeedphase}
\end{figure}

\subsection{Effects of stellar feedback}\label{stellarfeedback}

Stellar feedback is the main reason why the black holes do not grow significantly to produce observable flux \citep{Johnson2007}. Figure~\ref{halomass} indicates that the stellar mass comprises around 1\% of the host halo (virial) mass across the simulated redshift range, providing significant feedback in the interstellar medium (ISM). Starting from $z=15$, the virial temperature of our target halo remains 
above 10$^4$~K, often taken as the threshold for a more efficient mode of star formation \citep{Oh2002}. Thus, our target halo should be large enough to experience vigorous stellar feedback, at least locally \citep[e.g.][]{Jeon2015,Abe2021}. In addition, Fig.~\ref{temden} and \ref{metal} show the projected gas mass-averaged temperature and metallicity distribution in the larger-scale cosmological environment, as well as the halo hosting the target black hole. The plots indicate that the temperature in the halos and filamentary features of the cosmic web is generally high, implying that the gas is being heated by the stellar feedback. Furthermore, the metallicity in the virialized structures is significant, reaching levels of $\gtrsim 1\%$~solar, implying that past star formation and supernova feedback were common. While higher metallicity leads to more efficient cooling \citep[e.g.][]{Safranek2010}, the high temperatures in Figure~\ref{temden} imply that such cooling is sub-dominant compared to the stellar feedback heating.
The reason lies in the density dependence of the cooling and (photo-) heating rates, $\propto n^2$ and $n$, respectively, favouring the latter in the low- to intermediate-density environments in the bulk of the simulation box \citep[e.g.][]{Katz1996}.

To elucidate further, in Figure~\ref{temphase} we show temperature-density phase plots of the gas particles inside the target halo, at $z=15$, when the seed black hole is inserted, and at $z=5.61$,  when the simulation run ends. In both cases, a significant portion of the gas resides at a temperature of $10^4$\,K, corresponding to the temperature of photo-ionized gas affected by stellar feedback. The Bondi-Hoyle accretion is governed by the gas density and temperature near the black hole (see Equ.~\ref{bondi}). Since the gas in the vicinity of the black hole is heated by stellar feedback, resulting in higher temperature and lower density, black hole growth and accretion are insignificant under the conditions simulated here. We will discuss below, how accretion may have proceeded more efficiently in different physical circumstances (see Sect.~\ref{growth}). 

To explore how stellar feedback did affect black hole growth, we carried out simulation runs where star formation and stellar feedback will be turned off for 20~Myr for the gas particles that are swept by a spherical region around the target black hole. 
Such conditions are possible under strong LW backgrounds or dark matter-baryon streaming motions so that star formation is suppressed in a given region \citep[e.g.][]{Schauer2021}. The no-feedback zone's radius is 4 times the size of the black hole (softening-)kernel that contains 64 gas particles. Like the kernel, the region is determined on-the-fly and is updated at each timestep. However, by $z\sim15$, stars have already formed in the simulation, heating the existing gas in the halo. To self-consistently treat DCBH seeds, they should be inserted at earlier redshifts before stars have formed in the vicinity of the target halo, to realize the required near-primordial conditions for their formation \citep{Natarajan2017}. For the typical stellar seed black holes considered here, however, stars must have formed before the seeds, to provide the massive progenitors for any black hole remnant. 

In carrying out our no-feedback numerical experiment, we identify gas at much higher densities $n_{\rm H}\sim$10$^4$ cm$^{-3}$ in the halo compared to the original runs where the highest density of gas is $n_{\rm H}\sim$10$^2$ cm$^{-3}$ (Fig.~\ref{temphase}). 
However, the black hole does not remain in such density peaks but wanders away to low density regions (Fig.~\ref{nofeedphase}). This wandering behavior has been seen in previous simulations \citep[e.g.][]{Bellovary2019,Pfister2019,Ma2021} and observations \citep[e.g.][]{Reines2020,Mezcua2020} as well. 

Under such conditions, the overall black hole growth is even lower when compared to the original runs. This black hole `wandering' is likely a physical effect enhanced by numerical artifacts. Physically, a low-mass black hole may not be able to sink into a dense central region by dynamical friction as galaxies at high redshifts have irregular/clumpy morphologies with no well-defined centers \citep{Boldrini2020,Ma2021}. Numerically, 
the black hole is perturbed by gas and dark matter particles that are comparable or more massive than the black hole itself because of low resolution. Therefore, although we can produce dense regions around the black hole by suppressing star formation and stellar feedback, the black hole will not stay in those regions to efficiently accrete. In contrast, in our original runs with star formation and feedback included, dynamical friction from stellar particles and the feedback-induced lower gas density near the black hole allow the seeds to generally remain in the high density regions. Thus stellar feedback is the dominant cause of the inefficient black hole growth, at least for the case of light, stellar seeds. It is likely that in nature only more massive black hole seeds (in more massive galaxies with well-defined centers) can stay embedded in dense clouds to have significant accretion. 

Furthermore, the distance between the black hole and dense regions with $n_{\rm H}\sim$10$^4$ cm$^{-3}$ is typically $\gtrsim10^{2}$~pc. 
For comparison, the Bondi radius at this time is only $\sim0.1-10$\,pc, depending on the seed mass, as the black hole mass does not grow significantly.
If the black hole were to stay in the high-density regions and accrete all such nearby gas, it could have grown significantly more massive, possibly reaching $\sim 10^7$\,M$_{\odot}$ by the end of the simulation. However, in reality, such an occurrence is unlikely, as feedback from the black hole as it accretes will heat and hinder additional gas from falling in, and efficient mechanisms are needed to capture and maintain the black hole in the dense regions and bring the gas at $\sim100$\,pc to near the black hole to be accreted 
\citep{Milosavljevic2009,Hopkins2010,Davis2020}. Simulations with much higher resolution will be required to predict the dynamics of relatively low-mass black holes and their small-scale ($<10$ pc) environments.  

We note that our target halo represents a typical host system in the Universe at high redshifts, in terms of mass scale. It is evident from Figure~\ref{halomass} that our simulated host halo and stellar masses are significantly lower than what is observed with the \textit{JWST} in the Cosmic Evolution Early Release Science (CEERS) survey \citep{Finkelstein2022}. Moreover, all our seed black holes are less massive than previous quasar observations \citep{Woods2019,Inayoshi2020}, or recent CEERS AGN observations \citep{Kocevski2023,Larson2023}. As CEERS is not an extremely deep survey and past high-redshift quasar observations have been biased towards the brightest objects, these observations should detect the rarer and more extreme cases, whereas our target halo and central black holes represent a more common and typical case. If the stellar seed black holes formed in a rarer and more massive halo, the effects of stellar feedback will be reduced with a higher gravitational potential providing larger amounts of gas to the black hole \citep{Dimatteo2008}. This leads to a higher accretion rate, enhancing the black hole observability. Such effects of higher accretion are discussed in Section~\ref{growth}. Finally, we note that our target black holes do not obey the empirical $M-\sigma$ relation \citep{Evrard2008,Kormendy2013}, shown here as a reference point, other than at select redshifts. Given that this relation is expressing a population average, it is not clear how to connect it to the growth history of individual cases.

\section{JWST Observability}\label{4}

To assess observability, we convert the simulated black hole mass and accretion rate into its predicted flux observed at 2 \micron\ to compare with the sensitivity of the \textit{JWST} F200W filter. Following the equations and assumptions of \citet{Jeon2014}, we convert the mass accretion rate to total (bolometric) luminosity produced by the black hole via  
\begin{equation}
   L_{\rm BH}\simeq 6\times10^{39}~\text{erg s}^{-1}\frac{\dot{M}_{\rm BH}}{10^{-6}~\text{M}_\odot~\text{yr}^{-1}}\mbox{\ ,}
   \label{bhl}
\end{equation}
employing a radiative efficiency of $\epsilon_{\rm EM}\simeq 0.1$.
Assuming that every (stellar) black hole accretes with an Eddington ratio of $\sim10^{-2}$ (see Fig.~\ref{stellarseedratio}), the above expression can be written in terms of the black hole mass:
\begin{equation}
   L_{\rm BH}\simeq 10^{-2}L_{\rm Edd}\simeq1.7\times10^{39}~\text{erg s}^{-1}\left(\frac{M_{\rm BH}}{1000~\text{M}_\odot}\right)\mbox{\ ,}
   \label{bhl_eddratio}
\end{equation}
where $L_{\rm Edd}$ is the Eddington luminosity.

We decompose this total luminosity into a thermal multi-color disc (MCD) component at lower energies, including the wavelength that is redshifted to the 2\,\micron\ peak of the F200W filter, and a non-thermal power-law one at shorter wavelengths. We assume that the MCD and non-thermal components each contribute about half of the total luminosity \citep[e.g.][]{Kuhlen2005}. The specific luminosity of the MCD can be expressed as follows:
\begin{equation}
    L_{\rm MCD}(\nu,M_{\rm BH},\dot{M}) \propto A\nu^{1/3}\int_{x_{\rm in}}^{x_{\rm out}}\text{d} x \frac{x^{5/3}}{e^x-1}\mbox{\ ,}
    \label{mcd}
\end{equation}
where $A$ is a normalization factor. The integration limits, $x_{\rm in}$ and $x_{\rm out}$, depend on the frequency, $\nu$, as well as the black hole mass and accretion rate \citep{Jeon2014}. We integrate this spectrum up to 0.2 keV/$h_p$, taken as the highest frequency where the MCD luminosity dominates, with $h_p$ being Planck's constant. Equating the result to half the total luminosity, we determine $A$, which is thus a function of black hole mass and accretion rate.

We evaluate the normalized spectrum at $2/(1+z)$ \micron, corresponding to a frequency of $\nu_0(z) = 1.5\times10^{14}~\text{Hz}~(1+z)$. This emitted luminosity gives rise to an observed flux of
\begin{multline}
    f_{\nu, {\rm obs}}  \simeq 10^{-6}~\text{nJy}\left(\frac{L_{\rm MCD}(\nu_0(z),M_{\rm BH},\dot{M})}{10^{21}~\text{erg s}^{-1}~\text{Hz}^{-1}}\right) \\ \left(\frac{1+z}{10}\right)\left(\frac{D_L(z)}{100~\text{Gpc}}\right)^{-2}
    \label{fluxeq}
\end{multline}
Here $D_L(z)$ is the luminosity distance at redshift $z$, normalized to the value at $z=10$. To motivate the normalization for the specific source luminosity in the expression above, again assuming an Eddington ratio of $10^{-2}$ (see Fig.~\ref{stellarseedratio}), we can approximately write the MCD specific luminosity in terms of $M_{\rm BH}$ only:
\begin{equation}
\nonumber
       L_{\rm MCD}(\nu_0(z),M_{\rm BH}) \simeq 4.6\times10^{21}~\text{erg s}^{-1}~\text{Hz}^{-1}\left(\frac{ M_{\rm BH}}{1000 {\rm \,M}_\odot}\right)^{1.3}\mbox{\ ,}
\end{equation}
valid for $z\sim10$.

\begin{figure}
    \centering
    \includegraphics[width=0.5\textwidth]{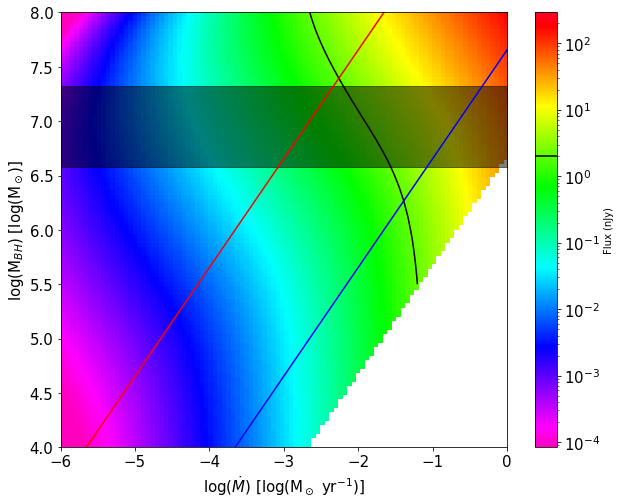}
    \caption{Predicted flux at 2~$\micron$ for a range of black hole masses and accretion rates, following Equations~\ref{mcd} and \ref{fluxeq}. In calculating the resulting fluxes, we assume the redshift inferred for the most distant AGN currently observed, $z=8.7$ \citep{Larson2023}. The bottom edge of the plot and white area below is the highly super-Eddington accretion regime ($\geq10\dot{M}_{\rm Edd}$) where the spectral model used may not be applicable. The black line marks the boundary where the predicted flux equals the \textit{JWST} NIRCam sensitivity for the F200W filter. Sources to the right of this line are predicted to be observable. The red line indicates, for a given black hole mass, the accretion rate that corresponds to an Eddington ratio of $10^{-2}$, as encountered in our simulations (see Fig.~\ref{stellarseedratio}), and the blue line the Eddington accretion rate for a given mass. The black shaded region reproduces the mass-range inferred for the AGN observed in \citet{Larson2023}, reflecting the uncertainties. Evidently, higher masses and accretion rates, compared to our simulated black holes (see Fig.~\ref{stellarseed}), are necessary to allow detection with the \textit{JWST}.}
    \label{predction}
\end{figure}

In the following, we compare the predicted flux of the target black holes with various initial masses to the sensitivity of the F200W filter at $\sim2$ nJy. This value corresponds to the 5$\sigma$ limiting AB magnitude of the CEERS survey at a magnitude of $\sim29$ \citep{Bagley2023}. In Figure~\ref{predction}, we show predicted fluxes (from Equ.~\ref{mcd} and \ref{fluxeq}) for a range of black hole masses and accretion rates at $z\simeq8.7$. This redshift corresponds to the currently highest redshift AGN detection \citep{Larson2023}. The sensitivity limit of the \textit{JWST} NIRCam F200W filter is indicated as a black line. The bottom edge and the white area below represent the highly super-Eddington accretion regime ($\geq10\dot{M}_{\rm Edd}$) where the spectral model used may not be applicable. Fig.~\ref{predction} also shows that the produced flux increases with the black hole mass for a given accretion rate. This trend is reflecting the harder MCD spectrum for less massive BHs with hotter disks (due to smaller inner radii), as a harder spectrum will have a lower near-infrared flux given the same luminosity/accretion rate. It is evident that our simulated stellar black hole seeds (see Fig.~\ref{stellarseed}), given their mass and accretion rates, do not reach the threshold for detectability. For an Eddington ratio of $10^{-2}$, as found for our simulated cases, black hole masses greater than $\sim10^{7.5}$\,M$_\odot$ are required to be observable at $z\sim9$ (see the red line in Fig.~\ref{predction}).

\begin{figure}
    \centering
    \includegraphics[width=0.5\textwidth]{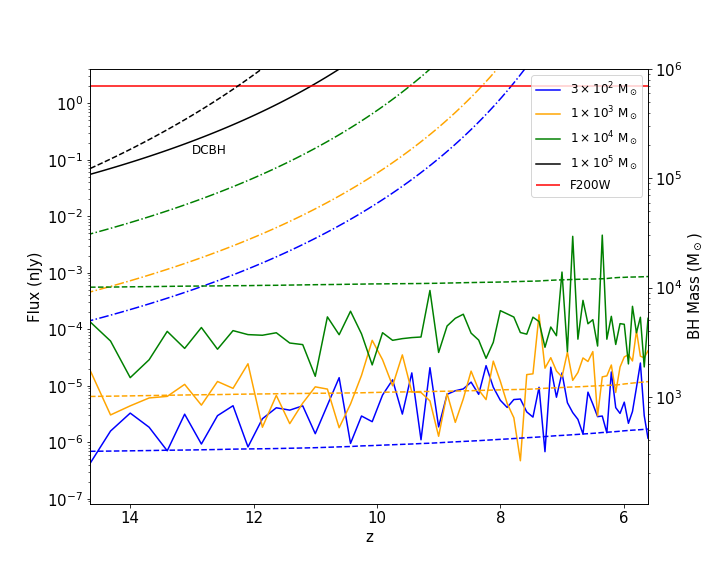}\par
    
    \caption{Predicted flux of the target black holes across redshifts, compared to the \textit{JWST} NIRCam sensitivity for the F200W filter (solid lines). We show the corresponding black hole masses across redshifts, as indicated on the right axis (dashed lines). For completeness, we also include the predicted flux for a DCBH with an initial mass of $10^5$ M$_\odot$ in the black solid line and its mass in the black dashed line, assuming that it follows the Eddington accretion rate. Although the accretion rates of the three simulated cases, as summarized in Fig~\ref{stellarseedratio}, are substantially sub-Eddington, it is useful to compare to the predicted fluxes, if the seed black holes were to follow Eddington accretion as well (dot-dashed lines). As can be seen, the stellar seed black holes do not produce high enough flux values to be detectable, unless a more optimistic accretion rate is realized.}
    \label{fluxseed}
\end{figure}

Due to the strong stellar feedback, accretion rates remain low, resulting in only very modest black hole growth. In Figure~\ref{fluxseed}, we show the corresponding fluxes from the black holes considered here. As is evident, the predicted fluxes remain below the NIRCam F200W filter sensitivity throughout, extending to $z\sim6-8$, around the time when reionization ended \citep{Fan2006,Finkelstein2019,Bosman2022}. While SMBHs with masses $\sim10^9$ M$_\odot$ are observed at high redshifts \citep{Wu2015,Banados2018}, the simulated growth of typical stellar seeded black holes in high-$z$ low-mass galaxies does not seem to be observable, unless more optimistic accretion scenarios are realized, as in the case of DCBH seeding and subsequent growth (Section~\ref{massiveseed}). 

Cosmological simulations can underestimate black hole accretion due to their low resolution \citep{Dimatteo2012,Schaye2015,Trinca2022}. This underestimation may have also occurred in our simulation with limited resolution. However, in this work, we are concerned with the behavior of a typical stellar seed black hole growing in a typical high-$z$ halo/galaxy rather than the rare and extreme growth settings. Moreover, previous higher resolution and smaller scale simulations of black hole accretion with sub-Eddington accretion rates generally showed steady accretion rates \citep{Jiang2019_2,Davis2020,Koudmani2022}, which is consistent with our average accretion behavior determined at lower resolution and larger scales, as shown in Fig.~\ref{stellarseed} and \ref{stellarseedratio}. Thus, our black hole growth and accretion are not unusual. The simulation results indicate that in typical environments around high-$z$ stellar seed black holes made of low-density, hot gas due to stellar feedback and chaotic dynamics, the Bondi accretion described by Equation~\ref{bondi} will not sufficiently boost the average stellar seed black hole masses to render them observable. If the SMBHs we do observe at high redshifts come from stellar seeds, such seeds must have accreted more efficiently closer to or above the Eddington rate. These higher accretion black holes may be observable, as discussed next.


\subsection{Massive seed model}\label{massiveseed}
We also consider growth for massive direct collapse black hole (DCBH) seeds. With mass around $\sim10^5$ M$_\odot$ \citep[e.g.][]{Becerra2018b,Becerra2018a}, inserting a DCBH seed in the simulation, in the same way as for the stellar seeds, would violate mass conservation more significantly as its mass is (about a factor of 10) larger than the typical mass of a gas cell in the simulation. A possible approach to avoid this problem would be to merge multiple gas particles with a combined mass corresponding to a DCBH into a single black hole sink particle. However, doing so would lower the central gas resolution, such that the gas conditions in the vicinity of the black hole, evaluated to estimate accretion parameters, would not be simulated at the same level as for the stellar seeds. We therefore here consider an idealized, analytical model of DCBH seed growth instead, and defer a proper numerical treatment to future work. Previous studies found that DCBHs with masses greater than $10^5$ M$_\odot$ could experience growth with Eddington ratios as large as 100 \citep{Pacucci2017,Basu2019}, efficiently accreting right from the start. We thus model DCBH growth assuming the Eddington accretion rate, in contrast to our stellar seed simulations with their Eddington ratios of $\sim10^{-2}$, as shown in Fig.~\ref{stellarseedratio}. Fig.~\ref{fluxseed} shows in black lines the flux (solid line) and mass (dashed line) of the DCBH model. The growth seeded by a DCBH would result in radiative fluxes that would be detectable by the \textit{JWST}, even as early as $z\sim11$, unlike the stellar seed scenarios. 

\subsection{Eddington and super-Eddington growth}\label{growth}

The results above do not exclude the possibility, however, that some low-mass seeds could sufficiently grow to allow detection. As shown in Fig.~\ref{fluxseed}, if the stellar seed black holes accreted at the Eddington rate, their resulting flux, indicated by the dot-dashed lines, would be observable even up to $z\sim10$, like the DCBH candidates in Section~\ref{massiveseed}. Observations exist of SMBHs with accretion rates close to or above Eddington \citep{Jin2016,Jin2017}. Furthermore, simulations with well resolved magnetohydrodynamics indicate that black hole accretion at super-Eddington rates is possible, where magnetic field or radiation-mediated viscosity causes instabilities that in turn drive material into the black hole \citep{Davis2020}. Moreover, non-uniform density around the black hole can create density waves in the accretion disk which can generate spiral shocks. These shocks apply Reynolds stress to the disk material, transferring angular momentum and driving accretion to extreme super-Eddington rates \citep{Jiang2014,Jiang2019}. Therefore, select SMBHs originating from stellar seeds may still be observable with the \textit{JWST} at high redshifts, under conditions that allow for Eddington or super-Eddington growth. It may be difficult to distinguish between the SMBHs formed from such stellar seeds accreting at extreme rates and DCBH seeds growing at rates close to Eddington. Modeling the signature of DCBH formation and subsequent growth \citep[e.g.][]{Dijkstra2016,Smith2017,Woods2019} will be needed to interpret upcoming high-redshift SMBH observations.

\subsection{Gravitational lensing}

Another avenue through which the less massive seeds could be observed is through gravitational lensing, a possibility that has been explored for Pop~III star clusters in the first galaxies \citep[e.g.][]{Zackrisson2015,Schauer2022}. Even if the SMBH seed does not accrete close to or above the Eddington rate, lensing magnification can enhance single point sources at high redshifts to be observable: If the source is located near the critical curve of the lens, magnification on the order of $10^3$ or larger is possible \citep{Welch2022}. In general, gravitational lensing amplifies the solid angle of the source on the sky, thus magnifying its observed flux by the same factor, given surface brightness conservation \citep{Jain2011}. 

\begin{figure}
    \centering
    \includegraphics[width=0.5\textwidth]{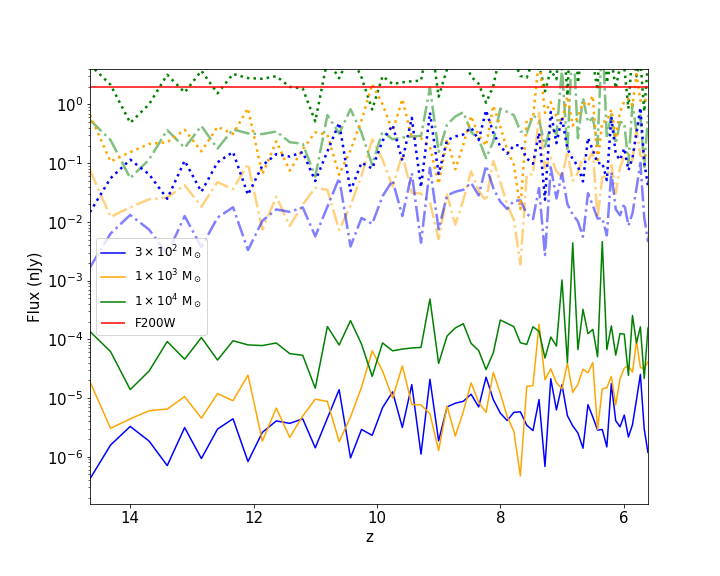}\par
    
    \caption{AGN flux levels magnified by gravitational lensing. The manner of presentation is the same as in Fig.~\ref{fluxseed}, except that now the dot-dashed lines represent the predicted flux for a lensing magnification of $\mu$=4000, and the dotted lines for $\mu$=35,000. These magnification values represent the lower and upper limits inferred for the lensing model applied to the high-redshift lensed star Earendel \citep{Welch2022}. For the lower magnification, the $10^4$ M$_\odot$ stellar seed would be observable with the \textit{JWST} at $z\lesssim7$. For the upper magnification limit, on the other hand, the $10^4$ M$_\odot$ seed would be observable even at $z>10$, whereas the $10^3$ M$_\odot$ seed could be detectable out to $z\sim7$.}
    \label{lensing}
\end{figure}

Fig.~\ref{lensing} illustrates the impact of such flux magnification for the observability of our seed black holes. We consider flux magnifications with the lower ($\mu$=4000) and upper ($\mu$=35,000) magnification estimates for the high-redshift lensed star-like object Earendel \citep{Welch2022}. With the lower magnification, the $10^4$ M$_\odot$ seed will be visible at $z\lesssim7$. With the larger magnification, the $10^4$ M$_\odot$ seed will be observable even at higher redshifts $(z>10)$. The $10^3$ M$_\odot$ seed will be observable as well near $z\sim7$. Thus, while the unique condition of an accreting seed black hole positioned near a critical curve of an astronomical lens is needed, detecting the seed black holes is possible even with the typical low accretion. \textit{JWST} has already observed a point-source star Earendel, and future observations may be able to identify further point-sources like it, possibly including seed black holes at high redshifts. A caveat of this approach is that while galaxies host millions of stars, they generally host only one central SMBH. The probability of a SMBH seed being located on the critical line for gravitational lensing is thus much lower than for a population of stars.

\section{Summary and Conclusions} \label{5}

We have run several cosmological simulations with \textsc{gizmo}, focusing on the growth of a single black hole in the center of a typical high-$z$ galaxy. Based on the simulated black hole accretion rate and mass over time, we calculate the predicted flux that would reach us today, comparing with the sensitivity of the \textit{JWST} F200W filter. Our results suggest that the typical stellar seed black holes will not grow sufficiently fast to reach the large masses required to be observable by $z\sim6-8$, even with \textit{JWST}, as black hole accretion is stunted by stellar feedback and the chaotic dynamics in high-$z$ low-mass galaxies. Our predicted near-constant, sub-Eddington accretion rates for the typical host galaxies at $z\gtrsim 8$ are found in previous studies as well \citep[e.g.][]{Jiang2019_2,Davis2020}. There is, however, the caveat of our limited numerical resolution, which does not allow us to represent all aspects of black hole growth, such as the radiation-hydrodynamical processes in the accretion disk.

We emphasize that even if the typical, stellar-seed black hole growth trajectories cannot be directly observed, a subset of them may sufficiently grow to become detectable under special conditions. Specifically, seeds could accrete at the Eddington or super-Eddington rates, which may be possible for stellar and massive black holes through magnetic fields, radiation viscosity, or shock instabilities \citep{Jiang2014,Jiang2019}. Even without high accretion, \textit{JWST} can observe extremely lensed point-sources at high redshifts \citep{Welch2022}. A seed black hole may be positioned near the critical curve of an astronomical lens like a galaxy cluster for its flux to be extremely magnified, and thus become observable.

There are other physical considerations that may affect high-redshift black hole observability, beyond the flux level produced by the accreting SMBH. Among them are the opacity of the accretion disk, the viewing angle, or the detailed spectral energy distribution (SED) set by the disk temperature. The AGN may be obscured by the material between us and the SMBH, such as the intergalactic medium or the host galaxy's interstellar medium \citep[e.g.][]{Smith2019,Gilli2022}, which will drive down the number of high-$z$ AGN that could be observed even further. 

Furthermore, we here have only considered AGN emission. In general, the flux from the AGN will be observed together with the stellar flux from the host galaxy. The currently highest-redshift AGN from CEERS/\textit{JWST} \citep{Larson2023} exhibits a subdominant emission compared to the stellar population. We thus encounter the additional challenge to unambiguously identify such a weak AGN. For the seed black hole growth history simulated here, approximately modeling the stellar emission with parameters for metal-poor (Population~II) stars, the AGN luminosity initially comprises a significant fraction of the total galaxy luminosity ($\sim 10\%$), becoming less important toward lower redshifts ($\sim 0.1\%$). This trend is a reflection of the limited growth of the seed black holes, whereas the halo stellar mass increases throughout. The signature of the stellar component is thus expected to become more prominent toward the lower end of the redshift range considered here \citep[e.g.][]{Agarwal2013}. A comprehensive modeling of the emerging SED is clearly needed to go beyond this idealized treatment \citep[e.g.][]{Natarajan2017,Nakajima2022}.

Finally, the number of AGN detectable in a given survey depends both on the flux limit and the survey area, while we have only addressed the first issue in this work. Estimating the survey area needed to detect low-luminosity AGN is beyond the scope of this work, as here we only focused on one massive halo and one SMBH instead of a population, with all cases considered here being too faint to be observable with the {\it JWST}. Future work examining a larger cosmological volume that includes the rare cases of high accretion or extreme gravitational lensing will require to constrain the survey area needed for low-luminosity AGN detections, for a given flux sensitivity.

The emerging picture of the co-evolution of stars and central black holes in the early Universe is complex, with a majority population of high-$z$ dwarf galaxies that likely host sub-dominant, weakly accreting AGN.
To account for the SMBHs observed in the first quasars at $z\gtrsim 7$, early black hole growth needs to be boosted, either with DCBH-like, heavy seeds, and/or more efficient (Eddington or super-Eddington) accretion channels. The branching ratio between such `normal' and accelerated assembly pathways is not known yet. With \textit{JWST} finally opening up the formative stages of primordial galaxy formation, and with the ideally complementary gravitational wave observatories that will become available in the next decade \citep[e.g.][]{Amaro2023}, progress in understanding the role of massive black holes in early cosmic evolution will be rapid and deep.


\section*{Acknowledgements}
The authors acknowledge the Texas Advanced Computing Center (TACC) at The University of Texas at Austin for providing HPC resources that have contributed to the research results reported within this paper. 

\section*{Data availability}
The data underlying this article will be shared on reasonable request to the corresponding author.



\bibliographystyle{mnras}
\bibliography{ms} 





\bsp	
\label{lastpage}
\end{document}